\documentclass[10pt]{article}
\usepackage[a4paper,left=1.7cm,right=1.7cm,top=1.65cm,bottom=1.75cm]{geometry}
\usepackage{amsmath,amssymb,mathtools,bm}
\usepackage{booktabs,array,longtable}
\usepackage{float}
\usepackage[font=small,labelfont=bf,skip=4pt]{caption}
\usepackage{enumitem}
\usepackage[compact]{titlesec}
\usepackage{tikz}
\usetikzlibrary{decorations.markings,decorations.pathmorphing}
\usepackage{microtype}
\usepackage{hyperref}
\usepackage{xcolor}
\hypersetup{colorlinks=true,linkcolor=blue!55!black,citecolor=blue!55!black,urlcolor=blue!55!black}
\allowdisplaybreaks
\setlist{nosep,leftmargin=1.8em}
\titleformat{\section}{\large\bfseries}{\thesection}{0.55em}{}
\titleformat{\subsection}{\normalsize\bfseries}{\thesubsection}{0.5em}{}
\titleformat{\subsubsection}{\normalsize\bfseries}{\thesubsubsection}{0.5em}{}
\titlespacing*{\section}{0pt}{1.25ex plus .4ex minus .2ex}{.55ex}
\titlespacing*{\subsection}{0pt}{1.0ex plus .3ex minus .2ex}{.35ex}
\titlespacing*{\subsubsection}{0pt}{.8ex plus .2ex minus .2ex}{.3ex}

\title{\vspace{-1.2cm}\bfseries About P-adic Numbers and Professor Igor Volovich on the Occasion
of his 80th Anniversary*}
\author{
Paul H. Frampton\\[1.5mm]
\small Dipartimento di Matematica e Fisica, Universit\`a del Salento, Lecce, Italy}
\date{Talk presented remotely to Moscow on October 5th, 2026}

\vspace{5mm}

\begin{document}
\maketitle

\vspace{1.0in}
\begin{abstract}
\noindent
We recall a 1987 paper by Volovich and our subsequent work with Okada
which established the Feynman rules for the p-adic string and the related non-local
effective field theory which found diverse physical applications. We add
some personal memories of visits to Moscow during 1989-1991.
\end{abstract}

\vfill

\noindent
*To appear in 2026 Proceedings of "New Horizons in Mathematical Physics",\\
published by the Steklov Mathematical Institute, Moscow.

\newpage

\noindent
I was delighted to accept this invitation to talk (remotely) at this Moscow conference
which celebrates the 80th anniversary of Professor Igor Volovich. My first
knowledge of Igor originated from the following paper:

\bigskip

\noindent
I.V. Volovich, {\it p-Adic String}, Class. Quant. Grav. {\bf 4,} L83 (1987).

\bigskip

\noindent
As background, I mention that I had previously collected and read
during 1968-74, some 500 papers about the 1968 Veneziano Model
(I did not reference all of them!) preparatory to writing my book {\it Dual Resonance Models (1974)}
and hence thought I knew all the ways
of rewriting the 4-point and N-point tree amplitudes.

\bigskip

\noindent
It was therefore a pleasant surprise to find 13 years later that Igor had discovered
a totally novel way of rewriting the 4-point amplitude using p-adic number
fields.

\bigskip

\noindent
The 4-point Veneziano model for external tachyons can be written
\begin{eqnarray}
A_4 & = & B(-\alpha(s), -\alpha(t)) + B(-\alpha(t), -\alpha(u)) \nonumber \\
 & & +B(-\alpha(u), -\alpha(s))
\label{A4}
 \end{eqnarray}
 where s,t,u are the Mandelstam variables, $\alpha(s) = 1 + s/2$ and $B(A,B) = \Gamma(A)\Gamma(B)/\Gamma(A+B)$ in which $\Gamma$ is the Euler gamma function.
 
\bigskip
\noindent
We may rewrite this as
\begin{equation}
A_4 = \int_{-\infty}^{\infty} dx |x|^{-\alpha(s)-1} |1-x|^{-\alpha(t)-1}
\label{A4prime}
\end{equation}
where the integral is over the real number field.

\bigskip

\noindent
The p-adic generalisation to $A_4^{(p)}$ for a given prime number p is
\begin{equation}
A_4^{(p)} = \int_{Q_p} dx |x|_p^{-\alpha(s)-1} |1-x|_p^{-\alpha(t)-1}
\label{Ap}
\end{equation}
where $|x|_p$ is the p-adic norm and $Q_p$ is the field of p-adic numbers.

\bigskip

\noindent
Using the properties of p-adic numbers which we shall not explain here
this amplitude can be rewriiten in  the multiplicative form
\begin{equation}
A_4^{(p)} = \Pi_{x=s,t,u} 
\left( \frac{1-p^{-\alpha(x)-1}}{1-p^{\alpha(x)}} \right)
\label{multiplicative}
\end{equation}

\bigskip

\noindent
Making use of the mass-shell condition
\begin{equation}
\alpha(s) + \alpha(t) + \alpha(u) = -1
\end{equation}
we can arrive at the additive form
\begin{equation}
A_4^{(p)} = \Sigma_x \frac{1-p^{-1}}{1-p^{\alpha(x)}}  - (2-p^{-1})
\label{additive}
\end{equation}
which has a simpler analytic structure than the real string..

\bigskip

\noindent
A fascinating formula is the adelic infinite product
\begin{equation}
\frac{1}{A_4}  \equiv \Pi_p A_4^{(p)}
\label{adelic}
\end{equation}
which shows how the real string is mathematically
composed from an infinite number of p-adic strings. 

\bigskip

\noindent
With my then new postdoc Yasuhiro Okada, we decided
to understand the p-adic N-point function and  the relevant Feynman
rules and worked intensively for a couple of weeks with the
conclusion that the correct propagator is
\begin{equation}
\frac{1-p^{-1}}{1-p^{\alpha(k^2)}}
\label{propagator}
\end{equation}
while the m-vertices $V_m$ are $V_3=1$ and for $m \geq 4$ they carry a factor
\begin{equation}
V_m = (-1)^{m+1} \Pi_{n=2}^{m-2}(n-p^{-1})
\label{vertex}
\end{equation}

\bigskip

\noindent
A natural next step was to derive an effective scalar field theory. The propagator
in Eq. (\ref{propagator}) inverts into a kinetic term
\begin{equation}
{\cal L}_{\phi}|_{kinetic} = \phi \left(  \frac{1-p^{1+\Box/2}}{1-p^{-1}} \right)  \phi
\label{kinetifc}
\end{equation}
in which $\Box$ is the D'Alembertian and so the kinetic term contains
an infinite number of derivatives and hence is non-local.

\newpage

\bigskip

\noindent
The two papers with Okada are:
P.H. Frampton and Y, Okada, Phys. Rev.Lett. {\bf 60,}  484 (1988); and
Phys.. Rev, {\bf D37,} 3077 (1988).

\bigskip

\noindent
The second paper, about the nonlocal effective field theory for the p-adic string,
was later used to study tachyon condensation in string theory and the
same theory has been used many times including in areas as diverse as black hole physics
and condensed matter theory. 

\bigskip

\noindent
Since this is a celebratory event, I'll spend my last five minutes on
selected memories of visiting Moscow in 1989, 1990 and 1991
(35-37 years ago) to discuss p-adic strings with Igor and his wife,
the physicist Irina Aref'eva.

\bigskip

\noindent
1989 was our first visit to Moscow.
For sight-seeing near Red Square, they wanted us to see the three
Cathedrals (Annunciation,Archangel and Dormition) in the Kremlin.
Tickets were sold miles away so, to save time, Irina suggested we act
as if a Russian schoolteacher with one of the visiting school groups
with the admonition "Don't say anything". This worked perfectly.

\bigskip

\noindent
In 1990, Igor and I took an overnight sleeper train from Moscow to
St. Petersburg, returning the following night. We visited the Kirov
Ballet, the Hermitage Museum and St. Petersburg's Steklov Institute 
where we met Ludwig Faddeev. Finally we went for tea with Irina's sister 
in a home with an extensive  library. While there I admired two volumes 
about Alexander Pushkin. When we left, Anna Aref'eva  insisted on 
presenting the books which I reluctantly accepted.

\bigskip

\noindent
When I arrived in Fenruary 1991, I was taken by Igor and Irina
from Moscow airport to Menezhnaaya Square where a 
pro-democracy demonstration was starting. When we arrived at
about 11 am there were less than a hundred protesters, outnumbered by
hundreds of armed soldiers encircling the square. There
were rumours that tanks were ready to move in and that hospital beds had 
been emptied for casualties. For the first hour, a repeat of the
Tiananmen Square massacre in China was a very scary possibility. By noon,
Mikhail Gorbachev had backed off.

\bigskip

\noindent
I hope that Igor continues to do
productive research for many more years to come.

\bigskip

\noindent
 Finally I list our two joint papers:

\bigskip

\noindent
\underline{\tt Cosmogenesis and primary quantization}\\
P. Frampton and I. Volovich.\\
Mod.Phys.Lett.{\bf A5,} 1825 (1990).

\bigskip

\noindent
\underline{\tt Wave function of the universe and p-adic gravity}\\
I. Aref'eva, B. Dragovich, P. Frampton and I. Volovich.\\
Int. J Mod. Phys. {\bf A6,} 4341 (1991).

\vspace{2.0in}

\noindent
Thank you for your attention.

\end{document}